\documentclass[twocolumn,showpacs,preprintnumbers,amssymb,nofootinbib]{revtex4}

\usepackage{graphicx}
\usepackage{bm} 

\begin{document}


\title{The black hole bomb and superradiant instabilities}

\author{Vitor Cardoso}
\email{vcardoso@teor.fis.uc.pt} \affiliation{Centro de F\'{\i}sica
Computacional, Universidade de Coimbra, P-3004-516 Coimbra,
Portugal}
\author{\'Oscar J. C. Dias}
\email{odias@ualg.pt} \affiliation{ Centro
Multidisciplinar de Astrof\'{\i}sica - CENTRA, Departamento de
F\'{\i}sica, F.C.T., Universidade do Algarve, Campus de Gambelas,
8005-139 Faro, Portugal}
\author{Jos\'e P. S. Lemos}
\email{lemos@kelvin.ist.utl.pt} \affiliation{ Centro
Multidisciplinar de Astrof\'{\i}sica - CENTRA, Departamento de
F\'{\i}sica, Instituto Superior T\'ecnico, Av. Rovisco Pais 1,
1049-001 Lisboa, Portugal}
\author{Shijun Yoshida}
\email{shijun@waseda.jp}
\affiliation{
Centro Multidisciplinar de Astrof\'{\i}sica - CENTRA,
Departamento de F\'{\i}sica, Instituto Superior T\'ecnico,
Av. Rovisco Pais 1, 1049-001 Lisboa, Portugal \footnote{Present address:
Science and Engineering, Waseda University, Okubo, Shinjuku, 
Tokyo 169-8555, Japan}}

\date{\today}

\begin{abstract}

A wave impinging on a Kerr black hole can be amplified as it scatters
off the hole if certain conditions are satisfied giving rise to
superradiant scattering. By placing a mirror around the black hole one
can make the system unstable.  This is the black hole bomb of Press
and Teukolsky. We investigate in detail this process and compute the
growing timescales and oscillation frequencies as a function of the
mirror's location. It is found that in order for the system black hole
plus mirror to become unstable there is a minimum distance at which
the mirror must be located.  We also give an explicit example showing
that such a bomb can be built.  In addition, our arguments enable us
to justify why large Kerr-AdS black holes are stable and small
Kerr-AdS black holes should be unstable.

\end{abstract}

\pacs{04.70.-s}

\maketitle
\newpage
\section{Introduction}
Superradiant scattering is known in quantum systems 
for a long time, after the problems raised by Klein's paradox \cite{manogue,greiner}. 
However, for classical systems  superradiant scattering 
was considered only much later in a paper by
Zel'dovich \cite{zel1} where it was examined what happens when scalar
waves hit a rotating cylindrical absorbing
object. Considering a wave of the form $e^{-i\omega t + i m \phi}$
incident upon such a rotating object, one concludes that if the
frequency $\omega$ of the incident wave satisfies
$\omega < m \Omega$,
where $\Omega$ is the angular velocity of the body, then the
scattered wave is amplified. 
It was also anticipated in \cite{zel1} that by surrounding the 
rotating cylinder by a mirror one could make the system unstable.

A Kerr black hole is one of the most interesting rotating objects for
superradiant phenomena, where the condition $\omega< m \Omega$ also
leads to superradiant scattering, with $\Omega$ being now the angular
velocity of the black hole \cite{bardeen,staro1,teu}.  Feeding back
the amplified scattered wave, one can extract as much rotational
energy as one likes from the black hole.  Indeed, if one surrounds the
black hole by a reflecting mirror, the wave will bounce back and forth
between the mirror and the black hole amplifying itself each
time. Then the total extracted energy should grow exponentially until
finally the radiation pressure destroys the mirror.  This is Press and
Teukolsky's black hole bomb \cite{press}. Nature sometimes provides
its own mirror. For example, if one considers a massive scalar field
with mass $\mu$ scattering off a Kerr black hole, then for $\omega <
\mu$ the mass $\mu$ effectively works as a mirror
\cite{damour,detweiler}.

Here we investigate in detail the black hole bomb by using a scalar
field model. Specifically, the black hole bomb consists of a Kerr
black hole surrounded by a mirror placed at a constant $r$, $r=r_0$, 
where $r$ is the Boyer-Lindquist radial coordinate.  We study the
oscillation frequencies and growing timescales as a function of the
mirror's location, and as a function of the black hole rotation.
A spacetime with a ``mirror'' naturally incorporated in it is
anti-de Sitter (AdS) spacetime, which has attracted a great deal
of attention recently. It could therefore be expected that
Kerr-AdS black holes would be unstable.  Fortunately, Hawking and
Reall \cite{hawking} have given a simple argument showing that, at
least large Kerr-AdS black holes are stable. As we shall show,
this is basically because superradiant modes are not excited for
these black holes. Furthermore, we suggest it is not only
possible but in fact highly likely that small Kerr-AdS black holes
are unstable.

\section{The black hole bomb}
\label{bhbomb}
\subsection{Formulation of the problem and basic equations}
We shall consider a massless scalar field in the vicinity of a
Kerr black hole, with an exterior geometry described by the line
element:
\begin{eqnarray}
ds^2 \!\!&=&\!\! -\left ( 1-\frac{2Mr}{\rho^2}\right )dt^2
-\frac{2Mr a\sin^2\theta}{\rho^2}\, 2 dt d\phi +\frac{\rho^2}{\Delta}\,dr^2 \nonumber \\
& &  +\rho^2 d\theta^2+\left ( r^2+a^2+\frac{2Mr
a^2\sin^2\theta}{\rho^2}\right )\sin^2\theta\, d\phi^2 \,, \nonumber \\
& &
 \label{metric}
\end{eqnarray}
with
\begin{eqnarray}
\Delta=r^2+a^2-2Mr\,, \qquad  \rho^2=r^2+a^2 \cos^2\theta \,.
 \label{metric parameters}
\end{eqnarray}
This metric describes the gravitational field of the Kerr black
hole, with mass $M$, angular momentum $J=M a$, and has an event
horizon at $r=r_+=M+\sqrt{M^2-a^2}$.
A characteristic and important parameter of a Kerr black hole is the angular velocity
of its event horizon $\Omega$ given by
\begin{equation}
\Omega=\frac{a}{2Mr_+}\,.
\label{Omega}
\end{equation} 
In absence of sources, which we consider to be our case, the
evolution of the scalar field is dictated by the 
Klein-Gordon equation in curved spacetime, $\nabla_{\mu}\nabla^{\mu}\Phi=0$. To make
the whole problem more tractable, it is convenient to separate the
field as \cite{brill}
\begin{equation}
\Phi(t,r,\theta,\phi)=e^{-i\omega t + i m \phi} S^m _l(\theta)
R(r)\,, \label{separation}
\end{equation}
where $S^m _l(\theta)$ are spheroidal angular functions, and the
azimuthal number $m$ takes on integer (positive or negative)
values. For our purposes, it is enough to consider positive
$\omega$'s in (\ref{separation}) \cite{bardeen}. Inserting this in
Klein-Gordon equation, we get the following angular and radial
wave equations for $R(r)$ and $S^m _l(\theta)$:
\begin{eqnarray}
& & \frac{1}{\sin \theta}\partial_{\theta}\left ( \sin \theta
\partial_{\theta} S^m _l \right )  \nonumber \\
& &\hspace{1cm}+
\left [  a^2 \omega^2 \cos^2
\theta-\frac{m^2}{\sin ^2{\theta}}+A_{lm} \right ]S^m _l =0\,,  
\label{wave eq separated1}
\\
& &  
\Delta\partial_r \left (
\Delta \partial_r R \right )+ {\bigl [} \omega^2(r^2+a^2)^2-2M a m
\omega r +a^2 m^2 \nonumber \\
& & \hspace{3.2cm} - \Delta (a^2\omega^2+ A_{lm}) {\bigr ]} R=0\,,
 \label{wave eq separated}
\end{eqnarray}
where $A_{lm}$ is the separation constant that allows the split of
the wave equation, and is found as an eigenvalue of
 (\ref{wave eq separated1}). For small $a \omega$, the regime we shall be interested
on in the next subsection, one has \cite{staro1,seidel}
\begin{eqnarray}
A_{lm}=l(l+1)+{\cal O}(a^2\omega^2)\,.
 \label{eigenvalues}
\end{eqnarray}
Near the boundaries of interest, which are the horizon, $r=r_+$,
and spatial infinity, $r=\infty$, the scalar field as given by
(\ref{separation}) behaves as
\begin{eqnarray}
\Phi \sim \frac{e^{-i\omega t}}{r}e^{\pm i\omega
r_*}\,\,,\,\,r\rightarrow \infty \label{bc1}
\\
\Phi \sim e^{-i\omega t}e^{\pm i(\omega-m\Omega)
r_*}\,\,,\,\,r\rightarrow r_+\,, \label{bc2}
\end{eqnarray}
where the tortoise $r_*$ coordinate is defined implicitly by
$\frac{dr_*}{dr}=\frac{r^2+a^2}{\Delta}$. 
Requiring ingoing waves at
the horizon, which is the physically acceptable solution, one must
impose a negative group velocity $ v_{\rm gr}$ for the wave
packet. Thus, we choose  $\Phi \sim
e^{-i\omega t}e^{- i(\omega-m\Omega) r_*}$.
However, notice that if
\begin{equation}
\omega < m \Omega \,, 
\label{super}
\end{equation}
the phase velocity$\,$
--$\frac{\omega-m\Omega}{\omega}$ will be positive.
Thus, in this superradiance regime, waves appear as
outgoing to an inertial observer at spatial infinity, 
and energy is in fact being extracted.  Notice that, since
we are working with positive $\omega$, superradiance will occur only
for positive $m$, i.e., for waves that are co-rotating with the black
hole. This follows from the time and angular dependence of the wave
function, $\Phi \sim e^{i(-\omega t+m\phi)}$. The phase velocity along the
angle $\phi$ is then $v_{\phi}=\frac{\omega}{m}$, which for 
$\omega>0$ and $m>0$ is positive, i.e., is in the same sense as the
angular velocity of the black hole.

Here we shall consider a Kerr black hole surrounded by a 
mirror placed at a constant Boyer-Lindquist radial $r$
coordinate with a radius $r_0$, so that the scalar field will be
required to vanish at the mirror's location, i.e., $\Phi(r=r_0)=0$.
With these two boundary conditions, ingoing waves at the horizon and a
vanishing field at the mirror, the problem is transformed into an
eigenvalue equation for $\omega$.  

The frequencies
satisfying both boundary conditions will be called Boxed Quasi-Normal
frequencies (BQN frequencies, $\omega_{BQN}$) and the associated modes
will accordingly be termed Boxed Quasi-Normal Modes (BQNMs). The quasi
stems from the fact that they are not stationary modes, and that BQN
frequencies are not real numbers. Instead they are complex quantities,
describing the decaying or amplification of the field.  One expects
that for a mirror located at large distances, or for small black
holes, the imaginary part of the BQNs will be negligibly small and thus the
modes will be stationary, corresponding to the pure normal modes of
the mirror in the absence of the black hole.
The BQNMs are of course different from the usual quasinormal modes
(QNMs) in asymptotically flat spacetimes, because the latter have no
mirror and satisfy outgoing wave boundary conditions near spatial
infinity, they describe the free oscillations of the black hole
spacetime.
In the following we shall compute these modes
analytically in a certain limit, and numerically by directly
integrating the radial equation (\ref{wave eq separated}).

\subsection{\label{sec:BH bomb analytical}The black hole bomb: analytical calculation of the unstable
modes}

In this section, we will compute
analytically, within some approximations, the unstable modes
of a scalar field in a black hole mirror system.
Due to the presence of a reflecting mirror
around the black hole, the scalar wave is successively impinging back on the 
black hole, and amplified.

We assume that $1/\omega \gg M$, i.e., that the Compton wavelength
of the scalar particle is much larger than the typical size
of the black hole. We will also assume, for simplicity, that $a\ll
M$. Following \cite{staro1,unruh}, 
we divide the space outside the event horizon in two regions,
namely, the near-region, $r-r_+ \ll 1/\omega$, and the far-region,
$r-r_+ \gg M$. We will solve the radial equation
(\ref{wave eq separated}) in each one of these two regions.
Then, we will match the near-region and the far-region solutions
in the overlapping region where $M \ll r-r_+ \ll 1/\omega$ is satisfied.
Finally, we will insert a mirror around the black hole, and
we will find the properties of the unstable modes.

\subsubsection{\label{sec:BH Near region}Near-region wave equation and solution}
In the near-region, $r-r_+ \ll 1/\omega$, 
the radial wave equation can be written as
\begin{eqnarray}
\Delta\partial_r \left ( \Delta\partial_r R \right )+ r_+^4
(\omega-m\Omega)^2\,R - l(l+1)\Delta\,R=0 \, .
 \label{near wave eq}
\end{eqnarray}
To find the analytical solution of this equation, one first
introduces a new radial coordinate,
\begin{eqnarray}
z=\frac{r-r_+}{r-r_-}\, , \qquad 0\leq z \leq 1\,,
 \label{new radial coordinate}
\end{eqnarray}
with the event horizon being at $z=0$. Then, one has $\Delta
\partial_r=(r_+-r_-)z\partial_z$, and the near-region radial wave
equation can be written as
\begin{eqnarray}
& & \hspace{-0.5cm} z(1\!-\!z)\partial_z^2 R+ (1\!-\!z)\partial_z
R+ \varpi^2
\frac{1\!-\!z}{z} R - \frac{l(l+1)}{1\!-\!z} R=0\,, \nonumber \\
& &
 \label{near wave eq with z}
\end{eqnarray}
where we have defined the superradiant factor
\begin{eqnarray}
\varpi \equiv(\omega-m\Omega)\frac{r_+^2}{r_+-r_-} \,.
 \label{superradiance factor}
\end{eqnarray}
Through the definition
\begin{eqnarray}
R=z^{i \,\varpi} (1-z)^{l+1}\,F \,,
 \label{hypergeometric function}
\end{eqnarray}
the near-region radial wave equation becomes
\begin{eqnarray}
& &  \hspace{-0.5cm} z(1\!-\!z)\partial_z^2 F+ {\biggl [} (1+i\,
2\varpi)-\left [ 1+2(l+1)+ i\, 2\varpi \right ]\,z {\biggr ]}
\partial_z F \nonumber \\
& & \hspace{1.5cm}- \left [ (l+1)^2+ i \,2\varpi (l+1)\right ]
F=0\,.
 \label{near wave hypergeometric}
\end{eqnarray}
This wave equation is a standard hypergeometric equation
\cite{abramowitz}, $z(1\!-\!z)\partial_z^2
F+[c-(a+b+1)z]\partial_z F-ab F=0$, with
\begin{eqnarray}
& & \hspace{-0.5cm} a=l+1+i\,2\varpi \,,  \qquad b=l+1 \,, \qquad
c=1+ i\,2\varpi \,, \nonumber \\
& &
 \label{hypergeometric parameters}
\end{eqnarray}
and its most general solution in the neighborhood of $z=0$ is $A\,
z^{1-c} F(a-c+1,b-c+1,2-c,z)+B\, F(a,b,c,z)$. Using
(\ref{hypergeometric function}), one finds that the most general
solution of the near-region equation is
\begin{eqnarray}
 \hspace{-0.5cm} R &=& A\, z^{-i\,\varpi}(1-z)^{l+1}
F(a-c+1,b-c+1,2-c,z)\nonumber \\
& & +B\,z^{i\,\varpi}(1-z)^{l+1} F(a,b,c,z) \,.
 \label{hypergeometric solution}
\end{eqnarray}
The first term represents an ingoing wave at the horizon $z=0$,
while the second term represents an outgoing wave at the horizon.
We are working at the classical level, so there can be no outgoing
flux across the horizon, and thus one sets $B=0$ in
(\ref{hypergeometric solution}). One is now interested in the
large $r$, $z\rightarrow 1$, behavior of the ingoing near-region
solution. To achieve this aim one uses the $z \rightarrow 1-z$
transformation law for the hypergeometric function
\cite{abramowitz},
\begin{eqnarray}
& \hspace{-2cm} F(a\!-\!c\!+\!1,b\!-\!c\!+\!1,2\!-\!c,z)=
(1\!-\!z)^{c-a-b}  & \nonumber \\
&\times
\frac{\Gamma(2-c)\Gamma(a+b-c)}{\Gamma(a-c+1)\Gamma(b-c+1)}
 \,F(1\!-\!a,1\!-\!b,c\!-\!a\!-\!b\!+\!1,1\!-\!z) & \nonumber \\
&  \hspace{-0.2cm}+
\frac{\Gamma(2-c)\Gamma(c-a-b)}{\Gamma(1-a)\Gamma(1-b)}
 \,F(a\!-\!c\!+\!1,b\!-\!c\!+\!1,-c\!+\!a\!+\!b\!+\!1,1\!-\!z), & \nonumber \\
&
 \label{transformation law}
\end{eqnarray}
and the property $F(a,b,c,0)=1$. Finally, noting that when
$r\rightarrow \infty$ one has $1-z\rightarrow (r_+-r_-)/r$, one
obtains the large $r$ behavior of the ingoing wave solution in the
near-region,
\begin{eqnarray}
R &\sim& A\,\Gamma(1-i\,2\varpi){\biggl [}
\frac{(r_+-r_-)^{-l}\,\Gamma(2l+1)}{\Gamma(l+1)\Gamma(l+1-i\,2\varpi)}\:
r^{l}\nonumber \\
& &
 +\frac{(r_+-r_-)^{l+1}\,\Gamma(-2l-1)}{\Gamma(-l)\Gamma(-l-i\,2\varpi)}\: r^{-l-1}
{\biggr ]}.
 \label{near field-large r}
\end{eqnarray}
\subsubsection{\label{sec:BH Far region}Far-region wave equation and solution}

In the far-region, $r-r_+ \gg M$, the effects induced by the black
hole can be neglected ($a\sim 0$, $M \sim 0$, $\Delta \sim r^2$)
and the radial wave equation reduces to the wave equation of a
massless scalar field of frequency $\omega$ and angular momentum
$l$ in a flat background,
\begin{eqnarray}
\partial_r^2 (r R)+ \left [ \omega^2-l(l+1)/r^2
\right ] (r R)=0\,.
 \label{far wave eq}
\end{eqnarray}
The most general solution of this equation is a linear combination
of Bessel functions \cite{abramowitz},
\begin{eqnarray}
R=r^{-1/2}\left [ \alpha J_{\,l+1/2}(\omega r)+ \beta
J_{-l-1/2}(\omega r)\right ]\,.
 \label{far field}
\end{eqnarray}
For large $r$ this solution can be written as \cite{abramowitz}
\begin{eqnarray}
& & R \sim \sqrt{\frac{2}{\pi \omega}}\frac{1}{r}{\biggl [} \alpha
\sin(\omega r-l\pi/2)+ \beta \cos(\omega r+l\pi/2){\biggr ]},\nonumber \\
& &
 \label{far field-large r}
\end{eqnarray}
while for small $r$ it reduces to \cite{abramowitz}
\begin{eqnarray}
R \sim \alpha\, \frac{(\omega/2)^{l+1/2}}{\Gamma(l+3/2)}\: r^{l} +
\beta\, \frac{(\omega/2)^{-l-1/2}}{\Gamma(-l+1/2)}\: r^{-l-1}.
 \label{far field-small r}
\end{eqnarray}

\subsubsection{\label{sec:BH Matching}Matching the near-region and the far-region solutions}

When $M \ll r-r_+ \ll 1/\omega$, the near-region solution and the
far-region solution overlap, and thus one can match the large $r$
near-region solution (\ref{near field-large r}) with the small $r$
far-region solution (\ref{far field-small r}). This matching
yields
\begin{eqnarray}
& & \hspace{-0.5cm} A= \frac{(r_+-r_-)^l}{\Gamma(l+3/2)}
\frac{\Gamma(l+1)}{\Gamma(2l+1)}
 \frac{\Gamma(l+1-i\,2\varpi)}{\Gamma(1-i\,2\varpi)}
  \left (\frac{\omega}{2} \right )^{l+1/2}\!\!\alpha ,\nonumber \\
& &
 \label{relation A-alpha}
\end{eqnarray}
\begin{eqnarray}
\frac{\beta}{\alpha} &=& \frac{\Gamma(-l+1/2)}{\Gamma(l+3/2)}
\frac{\Gamma(l+1)}{\Gamma(2l+1)}
 \frac{\Gamma(-2l-1)}{\Gamma(-l)}
 \frac{\Gamma(l+1-i\,2\varpi)}{\Gamma(-l-i\,2\varpi)}\nonumber \\
& &
  \times \left (\frac{\omega}{2} \right
  )^{2l+1}\!\!(r_+-r_-)^{2l+1}\,.
 \label{relation beta-alpha}
\end{eqnarray}
Using the property of the gamma function,
$\Gamma(1+x)=x\Gamma(x)$, one can show that
$\frac{\Gamma(-l+1/2)}{\Gamma(1/2)}=\frac{(-1)^l 2^l}{(2l-1)!!}$,
$\frac{\Gamma(l+3/2)}{\Gamma(1/2)}=\frac{(2l+1)!!}{2^{l+1}}$,
$\frac{\Gamma(-2l-1\!)}{\Gamma(-l)}=\frac{(-1)^{l+1}l!}{(2l+1)!}$
and
$\frac{\Gamma(l+1-i\,2\varpi)}{\Gamma(-l-i\,2\varpi)}=i\,(-1)^{l+1}2\varpi\prod_{k=1}^l
(k^2+4\varpi^2)$. Then, the matching condition
 (\ref{relation beta-alpha}) yields
\begin{eqnarray}
\frac{\beta}{\alpha} &=& i\, 2\varpi\, \frac{(-1)^l}{2l+1} \left (
\frac{l!}{(2l-1)!!} \right )^2
\frac{(r_+ -r_-)^{2l+1}}{(2l)! (2l+1)!} \nonumber \\
& & \times \, \left (  \prod_{k=1}^l (k^2+4\varpi^2) \right )
\omega^{2l+1}\,.
 \label{relation beta-alpha 2}
\end{eqnarray}

\subsubsection{\label{sec:BH Mirror}Mirror condition. Properties of the unstable modes}

If one puts a mirror near infinity at a radius $r=r_0$, the scalar
field must vanish at the mirror surface. Thus, setting
the radial field (\ref{far field}) to zero yields the
extra condition between the amplitudes $\alpha$ and $\beta$, and
the position of the mirror $r_0$,
\begin{eqnarray}
\frac{\beta}{\alpha}=- \frac{J_{l+1/2}(\omega r_0)}{J_{-l-1/2}(\omega r_0)} \,.
 \label{relation beta-alpha mirror}
\end{eqnarray}
This mirror condition together with the matching condition
 (\ref{relation beta-alpha 2}) yield a condition
 between the position of the mirror and the frequency of the scalar
 wave,
\begin{eqnarray}
& &  \hspace{-0.7cm}\frac{J_{l+1/2}(\omega r_0)}{J_{-l-1/2}(\omega r_0)}= 
i(-1)^{l+1} \,\varpi\, \frac{2}{2l+1}
\left ( \frac{l!}{(2l-1)!!} \right )^2 \nonumber \\
& & 
  \times \, \frac{(r_+ -r_-)^{2l+1}}{(2l)! (2l+1)!}\left ( \prod_{k=1}^l (k^2+4\varpi^2)
\right ) \omega^{2l+1} \,. 
 \label{mirror eigen}
\end{eqnarray}
The solution of (\ref{mirror eigen}) can be
found in the approximation that applies suitably to this problem,
namely, $\omega\ll 1$, and ${\rm Re}(\omega)\gg {\rm Im}(\omega)$.
For very small $\omega$, the r.h.s. of (\ref{mirror eigen}) is very small and can be
assumed to be zero in the first approximation for $\omega$. 
This yields 
\begin{equation}
J_{l+1/2}(\omega r_0)=0\,, 
\label{zerob}
\end{equation}
which has well studied (real) solutions \cite{abramowitz}.
We shall label the solutions of (\ref{zerob}) as $j_{l+1/2,n}$:
\begin{equation}
J_{l+1/2}(\omega r_0)=0 \Leftrightarrow \omega r_0=j_{l+1/2,n}\,,
\label{solzerob}
\end{equation}
where $n$ is a non-negative integer number.
We now assume that the complete solution to (\ref{mirror eigen}) can
be written as $\omega\sim j_{l+1/2,n}/r_0+i\tilde{\delta}/r_0$, where we
have inserted a small imaginary part proportional to
$\tilde{\delta}\ll 1$.
One then has, from (\ref{mirror eigen}) 
\begin{eqnarray}
& & \hspace{-1cm} \frac{J_{l+1/2}(j_{l+1/2,n}+i\tilde{\delta})}
{J_{-l-1/2}(j_{l+1/2,n}+i\tilde{\delta})}= i(-1)^{l+1} \,\varpi\, 
\left ( \frac{l!}{(2l-1)!!} \right )^2 \nonumber \\
& & \hspace{-0.5cm}
  \times \,\frac{2}{2l+1}\, \frac{(r_+ -r_-)^{2l+1}}{(2l)! (2l+1)!}\left ( \prod_{k=1}^l (k^2+4\varpi^2)
\right ) \omega^{2l+1} \,. 
 \label{mirror eigen2}
\end{eqnarray}
Now, we can use, for small $\tilde{\delta}$ the Taylor expansion of the l.h.s.,
\begin{equation}
\frac{J_{l+1/2}(j_{l+1/2,n}+i\tilde{\delta})}{J_{-l-1/2}(j_{l+1/2,n}+i\tilde{\delta})}
\sim i\,\tilde{\delta}\, \frac{J^{\prime}_{l+1/2}(j_{l+1/2,n})}{J_{-l-1/2}(j_{l+1/2,n})}
\label{app}
\end{equation}
The quantities $j_{l+1/2,n}\,,J^{\prime}_{l+1/2}(j_{l+1/2,n})\,,J_{-l-1/2}(j_{l+1/2,n})$
are tabulated in \cite{abramowitz}, and can also easily be extracted using Mathematica.
Here it is important to note that 
$J'_{l+1/2}(j_{l+1/2,n})$ and $(-1)^{l}J_{-l-1/2}(j_{l+1/2,n})$ always have the same sign. 
Furthermore, for large overtone $n$, $j_{l+1/2,n} \sim (n+l/2)\pi$.
The frequencies of the scalar wave that are allowed by
the presence of the mirror located at $r=r_0$ (BQN frequencies) are then
\begin{eqnarray}
\omega_{BQN} \simeq \frac{j_{l+1/2,n}}{r_0}+i\delta \,,
\label{mirror frequencies}
\end{eqnarray}
where $n$ is a non-negative integer number, labelling the mode overtone number.
For example, the fundamental mode corresponds to $n=0$.
In (\ref{mirror frequencies}), $\delta={\rm Im}[\omega_{BQN}]$ is obtained by substituting
(\ref{app}) in (\ref{mirror eigen2}),
\begin{eqnarray}
\delta \simeq  -\gamma  
\frac{(-1)^{l}J_{-l-1/2}(j_{l+1/2,n})}{J^{\prime}_{l+1/2}(j_{l+1/2,n})}
\frac{j_{l+1/2,n}/r_0-m\Omega}{r_0^{\:2(l+1)}}\,,
\label{mirror frequencies imaginary}
\end{eqnarray}
where 
\begin{eqnarray}
\gamma &\equiv &  
\left ( \frac{l!}{(2l-1)!!} \right )^2
\frac{r_+^2(r_+ -r_-)^{2l}}{(2l)! (2l+1)!} \nonumber \\ 
& & 
\times \,\frac{2}{2l+1} \left ( \prod_{k=1}^l (k^2+4\varpi^2)
\right )  \left [j_{l+1/2,n}\right ]^{2l+1}.
\label{mirror frequencies imaginary2}
\end{eqnarray}
Notice that $\delta$ is very small for large $r_0$ and thus satisfies
the conditions that go with the approximation used, ${\rm Re}(\omega)\gg {\rm Im}(\omega)$.
Equations (\ref{mirror frequencies}) and
(\ref{mirror frequencies imaginary}) constitute the main results of this section.
Two important features of this system, black hole plus mirror, can already be 
read from the equations above:
first, from equations (\ref{mirror frequencies}) and (\ref{mirror frequencies imaginary}) 
one has, 
\begin{equation}
\delta \propto -({\rm Re}[\omega_{BQN}]-m\Omega)\,.
\label{imag}
\end{equation}
Therefore,
$\delta>0$ for ${\rm Re}[\omega_{BQN}]<m \Omega$, and $\delta<0$ for
${\rm Re}[\omega_{BQN}]>m \Omega$. The scalar field $\Phi$ has the
time dependence $e^{-i\omega t}=e^{-i{\rm Re}(\omega) t}e^{\delta t}$,
which implies that for ${\rm Re}[\omega_{BQN}]<m \Omega$, the
amplitude of the field grows exponentially and the BQNM becomes
unstable, with a growth timescale given by $\tau=1/\delta$.
In this case, we see that the system behaves in fact as a bomb,
a black hole bomb.
Second, 
\begin{equation}
{\rm Re}[\omega_{BQN}] = \frac{j_{l+1/2,n}}{r_0} \,, 
\label{re}
\end{equation}
showing that the wave frequency is proportional to the
inverse of the mirror's radius.  As one decreases the
distance at which the mirror is located, the allowed wave
frequency increases, and there will thus be a critical radius
at which the BQN frequency no longer satisfies the superradiant
condition (\ref{super}).  
Notice also that ${\rm Re}[\omega_{BQN}]$ as given by (\ref{re}) 
is equal to the normal mode frequencies of a spherical
mirror in a flat spacetime \cite{landau}. 
In the next subsection, when the
numerical results will also be available, we will return to this
discussion.

\subsection{\label{sec:BH bomb numerical}The black hole bomb. Numerical
approach}

\subsubsection{Numerical procedure}
In the numerical calculation for determining oscillation
frequencies of the modes, we use the same function as that defined
by Teukolsky \cite{teu2} given by (see also \cite{hughes})
\begin{equation}
Y=(r^2+a^2)^{1/2}R\, .
\end{equation}
Then, Teukolsky equation becomes a canonical equation, given by
\begin{equation}
{d^2\over dr_*^2}Y+VY=0\,, \label{mT-eq}
\end{equation}
where
\begin{eqnarray}
V={K^2-\lambda\Delta\over(r^2+a^2)^2}-G^2-{d\over dr_*}G\,,
\end{eqnarray}
with $K=(r^2+a^2)\omega-am$, and $G=r\Delta(r^2+a^2)^{-2}$.
For the separation constant $\lambda=A_{lm}+a^2 \omega
^2-2am\omega$,  we make use of a well known series expansion in
$a\omega$, given by
\begin{eqnarray}
\lambda=a^2\omega^2-2am\omega+
\sum_{i=0}^{\infty}\,_0f^{lm}_i(a\omega)^i\,,
\end{eqnarray}
where $_0f^{lm}_i$ is the expansion coefficient (for the explicit
form, see, e.g., \cite{seidel}). In this study, we keep the terms 
in the expansion up to an order of $(a\omega)^2$. As mentioned, 
near the horizon, the physically acceptable solution of equation 
(\ref{mT-eq}) is the incoming wave solution, given by
\begin{eqnarray}
Y=e^{-i(\omega-m\Omega) r_*}[y_0+y_1 (r-r_+)+y_2 (r-r_+)^2+\cdots]\,,
\label{asyp-sol}
\end{eqnarray}
where $y_i$ is the expansion coefficient
determined by $\omega$ and $y_0$. Here, we do not show explicit
expression for the $y_i$'s because it is straightforward to derive it.

In order to obtain the proper solutions numerically, by using a
Runge-Kutta method, we start integrating the differential equation
 (\ref{mT-eq}) outward from $r=r_+(1+10^{-5})$ with the asymptotic
solution (\ref{asyp-sol}). We then stop the integration at the
radius of the mirror, $r=r_0$, and get the value of the wave
function at $r=r_0$, which is considered a function of the
frequency, $Y(r_0,\omega)$. If $Y(r_0,\omega)=0$, the solution
satisfies the boundary condition of perfect reflection due to
the mirror and the frequency $\omega$ is a BQN frequency, which we
label as $\omega _{BQN}$. In other words, the dispersion relation
of our problem is given by the equation $Y(r_0,\omega _{BQN})=0$.
In order to solve the algebraic equation $Y(r_0,\omega _{BQN})=0$
iteratively, we use a secant method. Here, it is important to note
that if the mode is stable or ${\rm Im}(\omega _{BQN}) <0$, the
asymptotic solution (\ref{asyp-sol}) diverges exponentially and
another independent solution, which is unphysical, damps
exponentially as $r_*\rightarrow -\infty$. 
\subsubsection{Numerical results}
\begin{figure}
\centerline{\includegraphics[width=7 cm,height=7 cm] {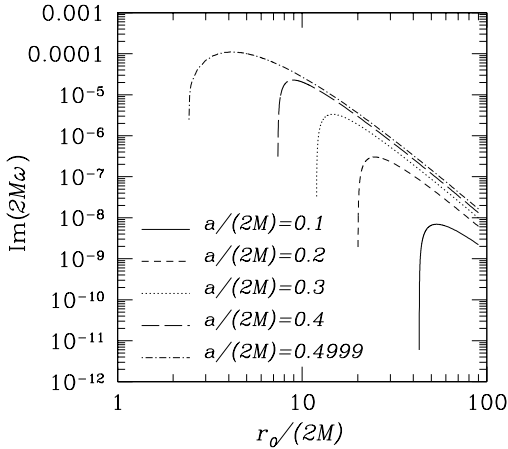}}
\caption{The imaginary part of the fundamental $(n=0)$ BQN frequency
($\omega_{BQN}$) as a function of the mirror's location $r_0$ is plotted.  
The plot refers to a $l=m=1$ wave. It is also
shown the dependence on the rotation parameter $a$.  One sees that for
$r_0$ greater than a critical value, $r_{0}^{\rm crit}$, depending
on $a$, there is the possibility of building a bomb.  Moreover, the
imaginary component of the BQN frequency decreases abruptly from its
maximum value to zero at $r_{0}^{\rm crit}$.  For $r_0<r_{0}^{\rm
crit}$ the BQNM is stable. Tracking the mode to yet smaller distances
shows that indeed it remains stable (the imaginary part of $\omega_{BQN}$ 
is negative).} 
\label{fig:BHBa}
\end{figure}

\begin{figure}
\centerline{\includegraphics[width=7 cm,height=7 cm] {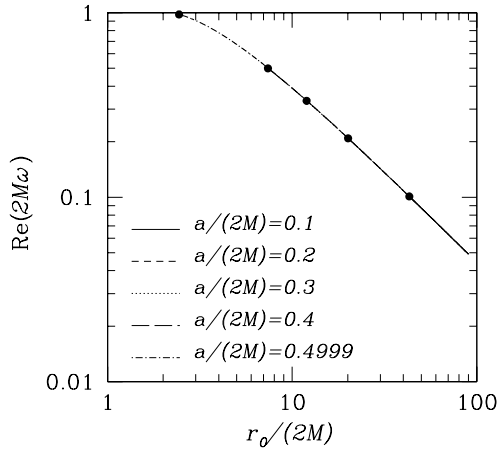}}
\caption{The real part of the fundamental $(n=0)$ BQN frequency
($\omega_{BQN}$) as a function of the mirror's location $r_0$ is
plotted.  The plot refers to a $l=m=1$ wave.
There is no perceptive $a$-dependence (as matter of fact
there is a very small $a$-dependence but too small to be noticeable).
Thus, the oscillation frequency basically depends only on $r_0$, and
for large $r_0$ goes as $1/r_0$, as predicted by the analytical
formula (\ref{mirror frequencies}). 
The dots indicate $r_{0}^{\rm crit}$ (cf. Fig. \ref{fig:BHBa}).}
\label{fig:BHBb}
\end{figure}

\begin{figure}
\centerline{\includegraphics[width=7 cm,height=7 cm] {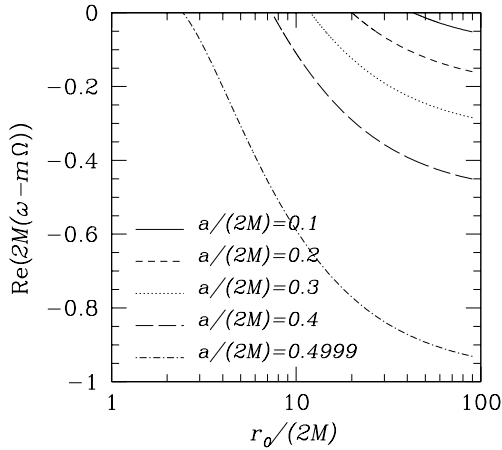}}
\caption{This figure helps in understanding why the instability
disappears for $r_0$ smaller than a certain critical value. The
condition for superradiance is $\omega-m\Omega <0$.
Since $\omega$ goes as $1/r_0$ (check Fig. \ref{fig:BHBb}), then it
is expected that the condition will stop to hold at a critical
$r_0$. This is clearly seen here. Note that the critical value
of $r_0$ is in excellent agreement with that shown in
Fig. \ref{fig:BHBa}.} \label{fig:BHBc}
\end{figure}

\begin{figure}
\centerline{\includegraphics[width=7 cm,height=7 cm] {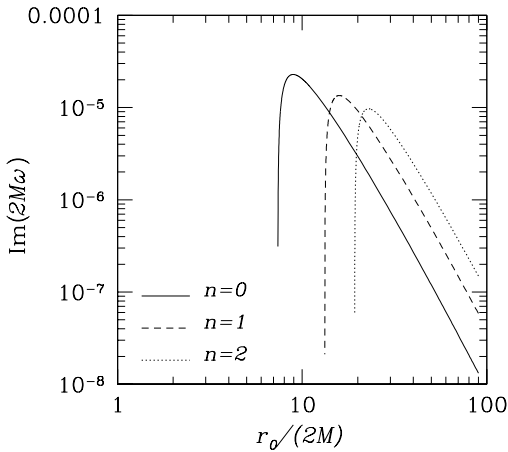}}
\caption{The imaginary part of the BQN frequency as a function of
$r_0$, for $a=0.4$ and for the three lowest overtones $n$, for
$l=m=1$. As expected from the general arguments presented, higher
overtones get stable at larger distances, and attain a smaller
maximum growing rate. } \label{fig:BHBd}
\end{figure}

\begin{figure}
\centerline{\includegraphics[width=7 cm,height=7 cm] {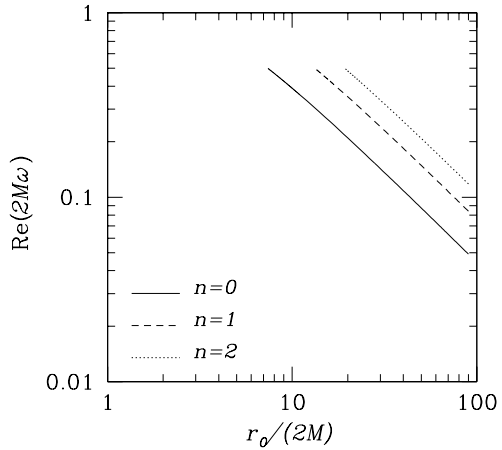}}
\caption{Same as Fig. \ref{fig:BHBd}, but for the real part of
$\omega_{BQN}$.} \label{fig:BHBe}
\end{figure}

\begin{figure}
\centerline{\includegraphics[width=7 cm,height=7 cm] {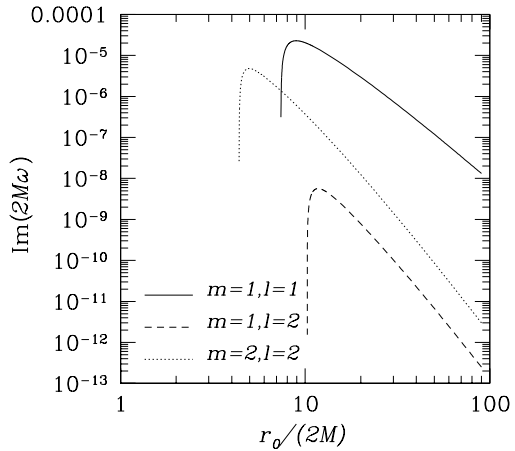}}
\caption{The imaginary part of the fundamental $\omega _{BQN}$ for an
$a=0.4$ black hole, as a function of the mirror's location $r_0$ here
shown for some values of $l,m$. Furthermore, as is evident from this figure and also
as could be anticipated, the larger $m$ the smaller $r_0$ can be,
still displaying instability.  Note however that the maximum
instability is larger for the $m=1$ mode. This is a general feature.
The imaginary part of the frequency seems to behave as ${\rm
Im}[\omega_{BQN}] \sim r_{0}^{-2(l+1)}$, which agrees with the
analytical prediction (\ref{mirror frequencies imaginary}).
} \label{fig:BHBf}
\end{figure}

\begin{figure}
\centerline{\includegraphics[width=7 cm,height=7 cm] {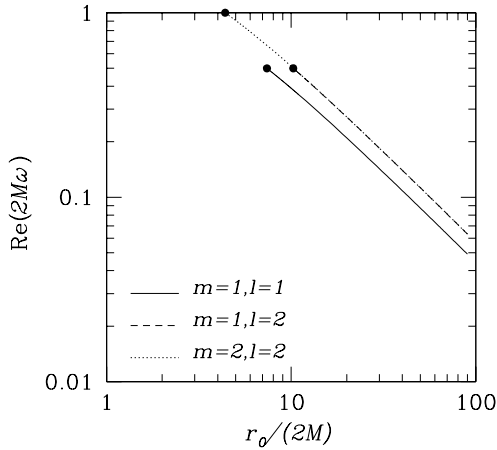}}
\caption{Same as Fig. \ref{fig:BHBf} but for the real part. Note that
there is an $l$-dependence of the real part of the frequency.  On the
other hand, there is no noticeable $m$-dependence, in accordance with
the analytical result (\ref{mirror frequencies}).}
\label{fig:BHBg}
\end{figure}

Our numerical results are summarized in
Figs. \ref{fig:BHBa}-\ref{fig:BHBg}. As we remarked earlier, we
only show the data corresponding to the unstable BQNMs.
We have also found the 
stable modes, but since they lead to no bomb we refrain from presenting them. 
>From the figures we confirm the analytical expectations. In addition we can 
discuss growing timescales, oscillation frequencies, energy extracted and efficiencies 
with great accuracy.
Figure \ref{fig:BHBa} plots the imaginary part of the BQN frequency for the
fundamental BQNM as a function of the mirror's location $r_0$ and 
the rotation parameter $a$. 
In figure \ref{fig:BHBb}, we show the real part of the BQN frequency for the
fundamental BQNM also as a function of $r_0$ and $a$. 
Supporting the analytical results, figure \ref{fig:BHBa}  shows that: 
(i) The instability is weaker (the
growing timescale $\tau=\frac{1}{{\rm Im}[\omega _{BQN}]}$ is larger)
for larger mirror radius, meaning that ${\rm Im}[\omega_{BQN}]$ decreases
as $r_0$ increases. This is also expected on physical grounds, as
was remarked in \cite{press}, if one views the
process as one of successive amplifications and reflections on the mirror. 
(ii) As one decreases $r_0$ the instability gets stronger, as
expected, but surprisingly, suddenly the BQNM is no longer
unstable. The imaginary component of $\omega_{BQN}$ drops from its
maximum value to zero, and the mode becomes stable at a critical
radius $r_0^{\rm crit}$.
Physically, this happens because superradiance generates wavelengths with
$\lambda > 1/\Omega$. So the mirror at a distance $r_0$ will ``see'' these wavelengths
if $r_0>r_0^{\rm crit}\sim \lambda\sim 1/\Omega$.
One can improve the estimate for $r_0^{\rm crit}$ using equations 
(\ref{super}), (\ref{imag}), and (\ref{re}), yielding
$r_0^{\rm crit} \sim \frac{j_{l+1/2,n}}{m\Omega}$.
This estimate for the critical radius matches very well
with our numerical data, even though the analytical calculation 
is a large wavelength approximation. 
In fact, to a great accuracy $r_0^{\rm crit}$ 
is given by the root of ${\rm Re}[\omega(r_0^{\rm crit})]-m\Omega=0$,
as is shown in figure \ref{fig:BHBc}.
Also supporting the analytical results, figure \ref{fig:BHBb}  shows that
${\rm Re}[\omega_{BQN}]$ behaves as $\frac{1}{r_0}$, which is consistent with
equation (\ref{mirror frequencies}). This means that it is
indeed the mirror which selects the allowed vibrating frequencies.
The results for higher mode number $n$ is shown in figures
\ref{fig:BHBd} and \ref{fig:BHBe}, and the behaviour agrees with the
picture provided by the analytical approximation.
The agreement between the analytical and numerical results is best seen in 
Table \ref{tab:comp}, where we show the lowest BQN frequencies obtained
using both methods.
In figures \ref{fig:BHBf} and \ref{fig:BHBg} we show the numerical results referring
to different values of the angular number $l$ and $m$.
Our numerical results indicate
that $\frac{1}{{\rm Im}[\omega _{BQN}]} \sim r_0^{-2(l+1)}$, in
agreement with the analytical result, equation (\ref{mirror
frequencies imaginary}). The numerical results also indicate that the
oscillating frequencies (${\rm Re}[\omega_{BQN}]$) scale with $l$,
more precisely, ${\rm Re[\omega_{BQN}]}$ behaves as ${\rm
Re[\omega_{BQN}]} \sim \pi/r_0(n+l/2)$. This behavior is also
predicted by the analytical study.

\vskip 1mm
\vskip 1mm
\vskip 1mm

\vskip 1mm
\vskip 1mm

\vskip 1mm
\vskip 1mm

\begin{table}
\caption{\label{tab:comp} The fundamental BQN frequencies
for a black hole with $a=0.8M$ and a mirror placed at $r_0=100M$.
The data corresponds to the $l=m$ modes, and the frequency is measured in units of 
the mass $M$ of the black hole. We present both
the numerical ($\omega_{BQN}^N$) and analytical ($\omega_{BQN}^A$)
results. Notice that the agreement between the two is very good, and
it gets better as $l$ increases. Also, we have checked that for very
large $r_0$ the two yield basically the same results.  }
\begin{ruledtabular}
\begin{tabular}{l|l|l}  \hline
\multicolumn{1}{c}{} &
\multicolumn{2}{c}{ $a=0.8M\,,r_0=100M$}\\ \hline
$l$ &$\omega_{BQN}^N$:&$\omega_{BQN}^A$:\\ \hline
1   &$8.75\times 10^{-2}+1.19\times 10^{-7}i$  &$8.99\times 10^{-2}+1.41\times 10^{-7}i$ \\ \hline 
2   &$1.13\times 10^{-1}+6.77\times 10^{-12}i$  &$1.15\times 10^{-1}+6.89\times 10^{-12}i$   \\ \hline 
3   &$1.37\times 10^{-1}+2.45\times 10^{-16}i$  &$1.39\times 10^{-1}+2.26\times 10^{-16}i$  \\ \hline  
\end{tabular}
\end{ruledtabular}
\end{table}
\vskip 1mm
Let us now take Press and Teukolsky example of a black hole with mass $M=1
M_{\odot}$ \cite{press}. We are now in a position to make a much
improved quantitative analysis.  We take $a=0.8\,M$, a large angular
momentum so that we make a full use of our results. In addition, to
better take advantage of the whole process, one should place the
mirror at a position near the point of maximum growing rate, but
farther. Thus, for the example, $r_0 \sim 22M \sim 33 \,{\rm Km}$ (see
Fig. \ref{fig:BHBa}).  This gives a growing timescale of about $\tau
\sim 0.6 \,{\rm s}$ (see also Fig. \ref{fig:BHBa}), which means that
every $0.6 \,{\rm s}$ the amplitude of the field gets approximately
doubled. This means that at the end of 13 seconds the initial
amplitude of the wave has grown to $10^7$ of its initial value, and
that thus the energy content is $10^{14}$ times greater than the
initial perturbation.  We consider there are no losses through the
mirror and assume the process to be adiabatic. Using the first law of
thermodynamics one can then set $\Delta M\sim \Omega \Delta J$, where
$\Delta M$ and $\Delta J$ are the changes in mass and angular momentum
of the black hole in this process, respectively.  Now, the black hole
is losing angular momentum in each superradiant scattering. Thus $a$
decreases. If we go to figure \ref{fig:BHBa} we see that $r_0^{\rm
crit}$ increases with decreasing $a$. At a certain stage $r_0^{\rm
crit}$ coincides with the position of the mirror at $r_0$, at which
point there is no more possibility of superradiance. The process is
finished. Thus from figure \ref{fig:BHBa}, or more accurately from our
numerics, one can find $\Delta\,a$, and thus $\Delta\,J$. Then
$\Delta\,M$ follows from $\Delta\,M\sim \Omega \Delta\,J$. In the
example this gives a total amount $\Delta\, M\sim 0.01 M$ of extracted
energy before the bomb stops functioning. The process has thus an
efficiency of 1\%, about the same order of magnitude as the efficiency
of nuclear fusion of hydrogen burning into helium ($\sim 0.7$\%).  If,
instead, the mirror is placed at $r_0 \sim 200 M \sim \, 300{\rm Km}$
one still gets a good growing timescale of about $15$ min.  This means
that at the end of $6$ hours the initial amplitude of the wave has
grown to $10^7$ of its initial value. In this case
$\Delta\,M \sim 0.1\,M$, with a 10\% efficiency, and one can show that the
efficiency grows with mirror radius $r_0$. Since the cost of mirror
construction scales as $r_0^2$, we see that small mirrors are more
effective. One can give other interesting examples.  For a black hole
at the center of a galaxy, with mass $M \sim 10^8 M_{\odot}$, $a=0.8
M$, and $r_0=22\,M$, the maximum growing timescale is of the order of
$2$ years.  Another interesting situation happens when the black hole
has a mass of the order of the Earth mass. In this case, by placing
the mirror at $r_0= 1\,{\rm m}$ one gets a growing timescale of about
$0.02 \, {\rm s}$. At the other end of the black hole spectrum one 
can consider Planck size black holes. 
\subsection{\label{sec:Zeldovich cylinder}Zel'dovich's cylinder surrounded by a mirror}
As a corollary, we discuss here electromagnetic
superradiance in the presence of a cylindrical rotating body, a situation
first discussed by Zel'dovich \cite{zel1}.
He noted that by surrounding this rotating body with a reflecting mirror one
could amplify the radiation, much as the black hole bomb process just
described.  Bekenstein and Schiffer \cite{bekenstein} have recently
elaborated on this.
An independent analytical approximation, similar in all respects to the one we
discussed earlier in the black hole bomb context, can also be applied here 
for finding the BQN frequencies of this system (conducting cylinder plus reflecting mirror),
and leads to almost the same results as for the black hole bomb.  
The imaginary component of $\omega_{BQN}$ is $\delta\propto -({\rm
Re}[\omega_{BQN}]-m\Omega)$.  The electromagnetic field has the time
dependence $e^{-i\omega t}=e^{-i{\rm Re}(\omega) t}e^{\delta t}$ and
thus, for ${\rm Re}[\omega_{BQN}]<m \Omega$, the amplitude of the
field grows exponentially with time and the mode becomes unstable,
with a growth timescale given by $\tau=1/\delta$. Second, ${\rm
Re}[\omega_{BQN}] \propto 1/r_0$, i.e., the wave frequency is
proportional to the inverse of the mirror's radius, as it was for the
black hole bomb.  Therefore, as one decreases the distance at which
the mirror is located, the allowed wave frequency increases, and again there
will be a critical radius at which the frequency no longer
satisfies the superradiant condition (\ref{super}).  If one tries to
use the system as it is, it would be almost impossible to observe superradiance in the
laboratory.  Take as an example a cylinder with a radius $R=10\, {\rm
cm}$, rotating at a frequency $\Omega=2\pi \times 10^2\, {\rm
s}^{-1}$, and a surrounding mirror with radius $r_0= 20 \, {\rm
cm}$. For the system to be unstable and experimentally detectable, the
minimum mirror radius is given by $r_{\rm crit} \sim
\frac{c}{m\Omega}$ (where we have reinstated the velocity of light
$c$), which yields $r_{\rm crit} \sim 1000 \,{\rm Km}$, for a $m=1$
wave. It seems impossible to use this apparatus to measure
superradiance experimentally.  A way out of this problem may be the
one suggested in \cite{bekenstein}: to surround the conducting mirror
by a material with a low velocity of light.  In this case the critical
radius would certainly decrease, although further investigation is
needed in order to ascertain what kind of material should be used.

\section{Are Kerr-AdS black holes unstable?}
A spacetime with a naturally incorporated mirror in it is anti-de
Sitter (AdS) spacetime, which has attracted a great deal of attention
recently due to the AdS/CFT correspondence and other matters.  As is
well known, anti-de Sitter (AdS) space behaves effectively as a box,
in other words, the AdS infinity works as a mirror wall.  Thus, one
might worry that Kerr-AdS black holes could behave as the black hole
bomb just described, and that they would be unstable. Hawking and
Reall \cite{hawking} have shown that, at least for large Kerr-AdS
black holes, this instability is not present.  The stability of large
Kerr-AdS black holes in four and higher dimensions can be understood
in yet another way, using the knowledge one acquired from the black
hole bomb study.  The black hole rotation is constrained to be
$a<\ell$, where $\ell$ is the AdS radius \cite{hawking}.
Large black holes are the ones for which $r_+>>\ell$. In
this case the angular velocity of the horizon $\Omega=
\frac{a}{r_+^2+a^2}(1-a^2/\ell^2)$ goes to zero and one expects that
the rotation plays a neglecting role in this regime, with the results
found for the non-rotating AdS black hole \cite{horo,cardoso} still
holding approximately when the rotation is non-zero.  The
characteristic quasinormal frequencies for large, non-rotating AdS
black holes were computed in \cite{horo,cardoso} showing that the real
part scales with $r_+$. Now, since $\Omega \rightarrow 0$ for large
black holes and the QNMs have a very large real part, one can
understand why there is no instability: superradiant modes are simply
not excited, as the condition for superradiance, $\omega<m\Omega$,
cannot be fulfilled.  We could try to evade this by going to higher
values of $m$, but then $l$ has also to be large ($l\geq m$). However,
for large $l$'s, the real part of the QNMs is known to scale with $l$
\cite{cardosol}, thus the condition for superradiance is never
fulfilled. 
What about small Kerr-AdS black holes,  $r_+<< \ell\,$? Considering 
the case $a$ small, $a<<r_+$ say, is enough for our purposes. 
In this situation the
horizon's angular velocity scales as the inverse of $r_+$,  and it
can be made arbitrarily large. 
Although the effect of rotation cannot be neglected in this case, 
the results of the QNM analysis for non-rotating AdS black holes
give some hints on what may happen.
For small non-rotating AdS black holes, the QN
frequencies have a real part that goes to a constant, independent of $r_+$, 
whereas the imaginary part goes to zero \cite{cardoso}.
If we add a small angular momentum per unit mass $a$ 
to the black hole, we do not expect
the real part of the QN frequency to grow significantly.
But, since $\Omega$ is very large anyway, the superradiance condition
$\omega<m\Omega$ will most likely be fulfilled.
Therefore we expect to be
possible to excite the superradiant instability in these
spacetimes.
\section{Conclusions}
To conclude, we have investigated the black hole bomb thoroughly, by
analytical means in the long wavelength limit, and numerically.  We
have provided both analytical and numerical accurate estimates for growing
timescales and oscillation frequencies of the corresponding unstable
Boxed Quasinormal Modes (BQNMs).  Both
results agree and yield consistent answers.  An important feature born
out in this work is that there is a minimum distance at which the
mirror must be located in order for the system to become unstable and
for the bomb to work.  Basically this is because the mirror selects
the frequencies that may be excited.  For distances smaller than this,
the system is stable and the perturbation dies off exponentially.
This minimum distance increases as the rotation parameter decreases.
We have given an explicit example where such a system works yielding a
reliable source of energy. By using appropriately this extracted energy 
one could perhaps build a black hole power plant. 
Although we have worked only with zero spin (scalar) waves,
we expect that the general features for other spins will be the
same. Moreover, it is known that a charged black hole, even in the
absence of rotation, provides a background for superradiance, as long
as the impinging wave is a bosonic charged wave (fermions do not
exhibit superradiance).  In this case, the critical
radius should be of order $r_{\rm crit}\sim \frac{1}{e\Phi}$, where
$e$ is the charge of the scalar particle and $\Phi$ is the black hole's
electromagnetic potential.  

We have also shown that a mirror
surrounding Zel'dovich's rotating cylinder leads to a system that
displays the same instabilities as the black hole bomb. However, for the 
instability to be triggered in an Earth based experiment, some improvments
must be made. In particular the cylinder should be surrounded by a material
with a low light velocity, since otherwise it would require huge mirror radius or huge
rotating frequencies. 

The study of the black hole bomb, and of the associated instabilities
allows one to better understand the absence of superradiance in large
Kerr-AdS black holes \cite{hawking} and moreover to expect that small
Kerr-AdS black holes will be unstable. 

Finally, it seems worth investigating whether or not this kind of
black hole bomb is possible in TeV-scale gravity.  In these scenarios,
one has four non-compact dimensions and $n$ extra compact
dimensions.  It might be possible that these extra compactified
dimensions work as a reflecting mirror, and therefore rotating black
holes in $4+n$ dimensions could turn out to be unstable.

We would also like to make a remark on a possible astrophysical 
application of this
black hole-reflecting wall system. It has been proposed in
\cite{putten}, and further discussed in \cite{aguirre}, that the
superradiant amplification process might provide the energy necessary
to feed the highly energetic gamma-ray burst. Magnetosonic plasma
waves, generated in the accreting plasma around an astrophysical black
hole, might enter in the waveguide cavity located between the
gravitational potential barrier of the black hole and the inner edge
of the accretion disk. Once there, the inner boundary of the accretion
disk might act as a reflecting wall, and waves can then suffer
multiple reflection and superradiant amplification, increasing their
energy.  This energy increase will continue until the magnetosphere
that surrounds the system is no longer capable of supporting the
energy pressure in the waveguide cavity. The wave energy would then be
released in a burst, and collimated into a relativistic jet by, e.g.,
the Blandford-Znajek process \cite{blandford}, and finally transferred
into the observed $\gamma$ -ray photons, as described by the Fireball
model \cite{piran}.
In order to better compare the energy and timescales derived from the 
superradiant model with the observational data taken from gamma-ray bursts,
it is appropriate to apply the methods of this paper to superradiant
cavities that have an accreting matter configuration of a torus or disk.

\vskip 2mm

\section*{Acknowledgements}
The authors acknowledge stimulating discussions with Ana  Sousa
and Jorge Dias de Deus, which have inspired us to do this work, and 
thank Emanuele Berti and Jacob Bekenstein for a critical reading of
the manuscript and for useful suggestions. 
This work was partially funded by Funda\c c\~ao para a Ci\^encia e
Tecnologia (FCT) -- Portugal through project CERN/FNU/43797/2001.
V.C. and O.J.C.D. acknowledge finantial support from FCT
through grant SFRH/BPD/2003. S.Y. acknowledges financial support
from FCT through project SAPIENS 36280/99.


\begin{thebibliography}{99}

\bibitem{manogue} For a very good review on the subject as well as a
careful explanation of various misinterpretations that have
appeared over the years concerning the Klein paradox, we refer the
reader to: C. A. Manogue, Annals of Physics {\bf 181}, 261 (1988).

\bibitem{greiner} W. Greiner, B. M\"uller and J. Rafelski,
{\it Quantum electrodynamics of strong fields}, (Springer-Verlag,
Berlin, 1985).

\bibitem{zel1} Ya. B. Zel'dovich,
Pis'ma Zh. Eksp. Teor. Fiz. {\bf 14}, 270 (1971)
[JETP Lett. {\bf 14}, 180 (1971)];
Zh. Eksp. Teor. Fiz {\bf 62}, 2076 (1972)
[Sov. Phys. JETP {\bf 35}, 1085 (1972)].


\bibitem{bardeen} J. M. Bardeen, W. H. Press and S. A. Teukolsky,
Astrophys. J. {\bf 178}, 347 (1972).

\bibitem{staro1} A. A. Starobinsky,
Zh. Eksp. Teor. Fiz {\bf 64}, 48 (1973)
[Sov. Phys. JETP {\bf 37}, 28 (1973)];
A. A. Starobinsky and S. M. Churilov,
Zh. Eksp. Teor. Fiz {\bf 65}, 3 (1973)
[Sov. Phys. JETP {\bf 38}, 1 (1973)].

\bibitem{teu} S. A. Teukolsky and W. H. Press,
Astrophys. J. {\bf 193}, 443 (1974).

\bibitem{press} W. H. Press and S. A. Teukolsky,
Nature {\bf 238}, 211 (1972).

\bibitem{damour} T. Damour, N. Deruelle and R. Ruffini,
Lett. Nuovo Cimento {\bf 15}, 257 (1976).

\bibitem{detweiler} S. Detweiler,
Phys. Rev. D {\bf 22}, 2323 (1980);
T. M. Zouros and D. M. Eardley,
Annals of Physics {\bf 118}, 139 (1979);
H. Furuhashi and Y. Nambu, gr-qc/0402037.

\bibitem{hawking} S. W. Hawking and H. S. Reall,
Phys. Rev. D {\bf 61}, 024014 (1999).

\bibitem{brill} D. R. Brill, P. L. Chrzanowski, C. M. Pereira, E. D. Fackerell and J. R. Ipser,
Phys. Rev. D {\bf 5}, 1913 (1972); S. A. Teukolsky, Phys. Rev.
Lett {\bf 29}, 1114 (1972).

\bibitem{seidel} E. Seidel,
Class. Quantum Grav. {\bf 6}, 1057 (1989).

\bibitem{unruh} W. G. Unruh,
Phys. Rev. D {\bf 14}, 3251 (1976).

\bibitem{abramowitz} M. Abramowitz and
A. Stegun, {\it Handbook of mathematical functions}, (Dover
Publications, New York, 1970).

\bibitem{landau} L. Landau and E. Lifshitz,
{\it Quantum mechanics, non-relativistic theory} 
(Mir, Moscow, 1974).

\bibitem{teu2} S.A. Teukolsky,
Astrophys. J. {\bf 185}, 635 (1973).

\bibitem{hughes} S. A. Hughes,
Phys. Rev. D {\bf 62}, 044029 (2000); Erratum-ibid. D {\bf 67},
089902 (2003).

\bibitem{leaver} E. W. Leaver,
Proc. R. Soc. London A{\bf 402}, 285 (1985).

\bibitem{bekenstein} J. D. Bekenstein and M. Schiffer,
Phys. Rev. D {\bf 58}, 064014 (1998).

\bibitem{horo} G. T. Horowitz, and V. Hubeny,
Phys. Rev. D {\bf 62}, 024027(2000); 
V. Cardoso and J. P. S. Lemos, Phys. Rev. D {\bf 64}, 084017 (2001); 
E. Berti, K. D. Kokkotas, Phys. Rev. D {\bf 67}, 064020 (2003).

\bibitem{cardoso} V. Cardoso, R. Konoplya, J. P. S. Lemos,
Phys. Rev. D {\bf 68}, 044024 (2003).

\bibitem{cardosol} V. Cardoso, unpublished.

\bibitem{putten} M. H. P. M. Putten,
Science {\bf 284}, 115 (1999).

\bibitem{aguirre} A. Aguirre, 
Astrophys. J. {\bf 529}, L9 (2000).

\bibitem{blandford} R. D. Blandford and R. L. Znajek,
Mon. Not. Roy. Astron. Soc. {\bf 179}, 433 (1997).

\bibitem{piran} T. Piran, 
Phys. Rep. {\bf 314}, 575 (1999);
P. M\'esz\'aros,
Ann. Rev. of Astron. and Astrophys. {\bf 40}, 137 (2002).


\end{thebibliography}
\end{document}